\documentclass[unsortedaddress,preprint,show pacs]{revtex4-1}
\usepackage{graphicx}
\usepackage{dcolumn}
\usepackage{amsmath}
\usepackage{amssymb}
\usepackage{amsfonts}
\usepackage{bm}
\usepackage{color}
\newcommand{\be}{\begin{equation}}
\newcommand{\ee}{\end{equation}}
\newcommand{\ba}{\begin{eqnarray}}
\newcommand{\ea}{\end{eqnarray}}
\def\diff{\mathop{\rm\mathstrut d\!}\nolimits}

\begin{document}
\title{On the determination of phase boundaries\\via thermodynamic integration across coexistence regions}
\author{Maria Concetta Abramo$^1$\footnote{Email: {\tt mcabramo@unime.it}}, Carlo Caccamo$^1$\footnote{Email: {\tt caccamo@unime.it}}, Dino Costa$^1$\footnote{Email: {\tt dcosta@unime.it}}, Paolo V. Giaquinta$^1$\footnote{Email: {\tt paolo.giaquinta@unime.it}}\\Gianpietro Malescio$^1$\footnote{Email: {\tt malescio@unime.it}}, Gianmarco Muna\`o$^1$\footnote{Email: {\tt gmunao@unime.it}}, and Santi Prestipino$^{1,2}$\footnote{Corresponding author. Email: {\tt sprestipino@unime.it}}}
\affiliation{$^1$Universit\`a degli Studi di Messina, Dipartimento di Fisica e di Scienze della Terra, Contrada Papardo, I-98166 Messina, Italy\\$^2$CNR-IPCF, Viale F. Stagno d'Alcontres 37, I-98158 Messina, Italy}
\date{\today}
\begin{abstract}
Specialized Monte Carlo methods are nowadays routinely employed, in combination with thermodynamic integration (TI), to locate phase boundaries of classical many-particle systems. This is especially useful for the fluid-solid transition, where a critical point does not exist and both phases may notoriously go deeply metastable. Using the Lennard-Jones model for demonstration, we hereby investigate on the alternate possibility of tracing reasonably accurate transition lines directly by integrating the pressure equation of state computed in a canonical-ensemble simulation with local moves. The recourse to this method would become a necessity when the stable crystal structure is not known. We show that, rather counterintuitively, metastability problems can be alleviated by reducing (rather than increasing) the size of the system. In particular, the location of liquid-vapor coexistence can exactly be predicted by just TI. On the contrary, TI badly fails in the solid-liquid region, where a better assessment (to within 10\% accuracy) of the coexistence pressure can be made by following the expansion, until melting, of the defective solid which has previously emerged from the decay of the metastable liquid.
\end{abstract}
\maketitle

\section{Introduction}

A recurrent theme in the numerical simulation of condensed-matter systems is that of metastability, which seriously plagues the determination of phase boundaries in all cases where the transition is accompanied by a spontaneous symmetry breaking. Metastability is evidenced in the dramatic slowing down of the system relaxation dynamics (be it true or fake, as in a Monte Carlo (MC) simulation), causing a noticeable shift of the transition point which prevents the correct identification of the system structure throughout the transition region. For example, a better control of coexistence conditions may be useful in the  design of complex-fluid systems with prescribed self-assembly properties~\cite{Whitesides}.

In order to overcome the metastability bottleneck a toolbox of smart methods has been developed through the years, which has made it unnecessary to wait for the spontaneous nucleation of the stable phase (say, solid) from the parental phase (liquid). As a matter of fact, only deep in the solid region the size of the critical nucleus is reduced to such an extent that the spontaneous formation of a solid embryo occurs within typical simulation times (once an embryo has formed, the growth of the solid from the liquid is very fast). Among the numerical methods which enable one to draw a melting line ``exactly'' ({\it i.e.}, with a negligible statistical error) we can at least mention the Einstein-crystal (or Frenkel-Ladd) method (see, {\it e.g.}, Ref.~\cite{Frenkel}), the MC phase-switch method~\cite{Wilding}, the interface-velocity method~\cite{ZykovaTiman}, and the interface-pinning method~\cite{Pedersen}. Another method which does not require an interface to form between coexisting phases is the Gibbs-ensemble method~\cite{Panagiotopoulos}, which, however, only works for the equilibria between fluid phases. While the three last mentioned algorithms allow one to directly compute coexistence parameters, a Frenkel-Ladd simulation rather aims at determining the free energy of the system in a reference crystalline state far from coexistence. The crystal free energy in any other state (either stable or metastable) is then obtained by resorting to thermodynamic integration (TI) along any path joining the given state to the reference state. The shortcoming shared by all these approaches is the necessity to assume {\it a priori} the knowledge of the crystalline structure, whose optimization may in fact be a rather daunting task if the system of interest is sufficiently complex (see, {\it e.g.}, the cases analyzed in Refs.\,\cite{Prestipino1,Abramo}). In such a case, metadynamics~\cite{Martonak} or a genetic algorithm~\cite{Oganov} may be useful ways out.

However, suppose that we do not have any of these powerful machinery at hand. If we attempt a stepwise approach to the solid-liquid transition from the liquid side ({\it e.g.}, by slowly increasing the density in an isothermal simulation with local moves) we invariably end up with the formation of a metastable liquid, thus bypassing the coexistence region. Eventually, the overcompressed liquid undergoes freezing into a more or less ordered solid and the pressure abruptly drops down. Further compressions lead to a regular increase of the pressure of the by-now solid system. Clearly, the resulting equation of state (which superficially recalls the van der Waals loop of mean-field theories) is not by itself sufficient to extract coexistence parameters (for example, the coexistence pressure may be wrong by 50\% or more, see below), and this would be the rule for numerical experiments implementing local particle moves only. Obviously, an explicit two-phase simulation where the position of the interface is recorded as a function of time~\cite{ZykovaTiman} would be a more straightforward and potentially very accurate method for locating solid-liquid coexistence but, in order to implement it, the crystalline structure would be required as an input. Based on such a discussion the question naturally arises as to which simulation features mostly affect the error made in estimating phase thresholds by exclusive resort to TI and, in particular, whether the simulation setup can be managed in such a way that the system relaxation time becomes acceptably small. Being able to answer these questions may greatly help in locating solid-liquid coexistence when the crystalline structure is unknown.

Similar problems would also be encountered when dealing with the condensation of vapor into liquid by isothermal compression. We know that, owing to the finiteness of the system and to the use of periodic boundary conditions, the heterogeneous fluid sample undergoes a sequence of geometric transitions inside the coexistence region~\cite{Furukawa,Binder1,Neuhaus,MacDowell1,MacDowell2,Binder2}. Such pseudo-transitions produce a series of jumps and plateaus in the pressure and chemical-potential equations of state, raising doubts on the possibility of accurately computing the chemical potential of the bulk liquid by moving across the two-phase region rather than circumventing the critical point. In fact, the grand-canonical simulations with reweighting performed by MacDowell {\em et al.} have already shown that the location of the condensation transition is nearly independent on the system size (see Fig.\,4 of Ref.\,\cite{MacDowell1}), but it is not clear whether standard canonical molecular dynamics (MD) or MC simulations would be accurate as well.  

In order to find the best strategy for making a reliable estimate of coexistence thresholds in an ordinary simulation with local moves we have considered the Lennard-Jones (LJ) model in two variants. Using periodic boundary conditions, we have simulated different system sizes and box shapes, eventually identifying a workable protocol to minimize the error due to the occurrence of metastability. Surprisingly, we found that too large a system size is detrimental to the accuracy of the chemical-potential reconstruction, for both the liquid-vapor and solid-liquid transitions. We expect that the suggestions coming from the analysis of LJ-type models will in fact be of more general use, thus allowing one to draw a reasonable melting line also in those cases where no other route can be successfully pursued.

This paper is organized as follows. In Sec.\,II we introduce the models under study and describe the simulation method in detail. Canonical-ensemble results for both the liquid-vapor and the solid-liquid transitions are discussed in Sec.\,III. A few concluding remarks are presented in the final section.

\section{Model and method}
\setcounter{equation}{0}
\renewcommand{\theequation}{2.\arabic{equation}}

In order to assess the ability of traditional simulation methods to predict accurate transition boundaries, any model fluid whose behavior is known with very high precision would suffice. Hence, we have chosen the paradigmatic LJ model. We have considered two variants of the LJ potential: (Model A) the original potential, truncated at $5\sigma$ and augmented with energy and pressure long-range corrections (whose critical and triple-point temperatures are respectively $T_{\rm c}\simeq 1.32\,\epsilon/k_{\rm B}$~\cite{Smit} and $T_{\rm t}\simeq 0.69\,\epsilon/k_{\it B}$~\cite{Mastny}, where $k_{\it B}$ is Boltzmann's constant); and (model B) the cut-and-shifted LJ potential,
\ba
u(r)=\left\{
\begin{array}{rl}
4\epsilon\left[(\sigma/r)^{12}-(\sigma/r)^6\right]-c\,, & \,\,\,{\rm for}\,\,r<r_{\rm cut}\\
0\,, & \,\,\,{\rm for}\,\,r>r_{\rm cut}
\end{array}
\,\,\,\,\,\,{\rm with}\,\,\,c=4\epsilon\left[\left(\frac{\sigma}{r_{\rm cut}}\right)^{12}-\left(\frac{\sigma}{r_{\rm cut}}\right)^6\right]\,,\right.
\nonumber \\
\label{2-1}
\ea
with $r_{\rm cut}=2.5\sigma$; no long-range corrections are required in this case. The critical temperature of model B is slightly less than $1.10\,\epsilon/k_{\rm B}$~\cite{Smit} (from now on, all quantities will be expressed in the units set by $\epsilon$ and $\sigma$).

As far as model A is concerned, we carried out canonical-ensemble, $NVT$ molecular-dynamics simulations (where $N$ is particle number and $V$ is volume) using the {\tt MOLDY} code~\cite{MOLDY} (in this package the temperature $T$ is set by a Nos\'e-Poincar\'e thermostat~\cite{Bond}). The sample consisted of $N=1372$ particles enclosed in a cubic simulation box whose edge varies according to the value of the number density $\rho=N/V$. Periodic boundary conditions (PBC) were applied. For model A, all simulation runs were started from scratch, {\it i.e.}, from an initial face-centered cubic (fcc) configuration with random momenta. The time length of each run was typically $25000$\,ps, with an integration time step of $5\times 10^{-15}$\,s; statistical averages were computed over the last $5000$\,ps only.

As for model B, we performed $NVT$ Metropolis MC simulations for samples of various sizes ($N=108,256,500,1372,4000$) in a cubic box with PBC. We also considered a system of 1500 particles in an elongated, cuboidal box. Simulation runs for model B were made in a sequence: for each state point along a path, the initial configuration was taken to be the last (rescaled) configuration generated at the previous state. For each state, a number $M$ of MC cycles (one cycle $=N$ elementary particle moves) were first produced to achieve equilibration, followed by other $M$ cycles over which the equilibrium averages were computed. All of our runs within the liquid-vapor region were made of as many as 4 million cycles ($M=10^7$ for $N=1372$), whereas we took $M=5\times 10^5$ ($M=10^6$ for $N=4000$) for higher densities. Statistical errors were estimated assuming no correlation between block averages (the equilibrium trajectory was usually divided in ten blocks).

Along an isothermal path, the excess Helmholtz free energy per particle was computed through the equation
\be
\beta f_{\rm ex}(\rho,T)=\beta f_{\rm ex}(\rho_1,T)+\int_{\rho_1}^{\rho}\left(\frac{\beta P(\rho',T)}{\rho'}-1\right)\frac{\diff \rho'}{\rho'}\,,
\label{2-2}
\ee
where $\beta=(k_{\rm B}T)^{-1}$. In order to set the free-energy offset, a low-density and high-temperature fluid was taken for reference. At this point, the chemical potential was computed by Widom's method~\cite{Widom}. The grid spacing along an isothermal path was typically $\Delta\rho=0.01$. The raw data were interpolated by spline functions and then integrated through Eq.\,(\ref{2-2}). For an integration along an isochoric path, a different formula was used:
\be
\beta f_{\rm ex}(\rho,T)=\beta_1f_{\rm ex}(\rho,T_1)-\int_{T_1}^{T}\frac{e(\rho,T')-(3/2)k_{\rm B}T'}{k_{\rm B}T'^2}{\rm d}T'\,,
\label{2-3}
\ee
where $e$ is the energy per particle. The grid spacing along isochoric paths was $\Delta T=0.05$. From the knowledge of $f_{\rm ex}$ one promptly derives the chemical potential $\mu$ and can thus identify a first-order transition point as the point where, {\it e.g.}, the $\mu(P)$ curves of the two phases cross each other at constant temperature. 

For a thorough check of our method, we computed the free energy of model B at a reference fcc-solid state by the Frenkel-Ladd method. This calculation provided a benchmark estimate of solid-liquid coexistence with which to compare.

\section{Results}
\setcounter{equation}{0}
\renewcommand{\theequation}{3.\arabic{equation}}

We carried out extensive simulations of models A and B, for various temperatures and with different simulation protocols, in order to gain as much information as possible on the ability of traditional simulation methods to predict accurate liquid-vapor and solid-liquid boundaries. In the following we keep the analysis of the two transitions distinct, since they have specificities which recommend a separate treatment.

\subsection{The condensation transition and the shape of the liquid-vapor interface}

Suppose we start with a stable vapor at a certain temperature $T<T_{\rm c}$ and then slowly increase the pressure $P$ until condensation occurs. Clearly, this simple-minded approach to liquid-vapor coexistence is doomed to fail since the vapor usually goes metastable. Quite different would be the outcome of the experiment if the system {\em density} $\rho$ (rather than its pressure) is increased in steps: in this case the path goes through the liquid-vapor region and the pressure is a continuous function of the density. The question is: will liquid-vapor coexistence be properly characterized by plain TI or, in other words, how correct is the chemical potential $\mu$ of the liquid as computed via Eq.\,(\ref{2-2})?

As remarked by Binder and coworkers in a number of recent papers~\cite{Binder1,MacDowell1,MacDowell2,Binder2}, but actually known since the early times of the computer-simulation era~\cite{Mayer}, a finite-size vapor system in a periodic simulation box undergoes, in the two-phase region, a sequence of so-called ``geometric transitions'', which are morphological transitions of the interface between liquid and vapor. For each geometric transition (actually a more or less pronounced crossover, depending on the system size $N$) $P(\rho)$ exhibits a drop, while staying roughly constant in the density interval between one pressure ``jump'' and the next one (we emphasize that these features of $P(\rho)$ are {\em equilibrium} characteristics elicited by the use of PBC). As $N$ grows, the jumps reduce in extent until just one perfect plateau only is left in the thermodynamic limit, extending from $\rho_{\rm v}$ to $\rho_{\rm l}$ (the bulk coexistence densities of vapor and liquid). While the density location of each geometric transition is only slightly size-dependent, the pressure level of each intermediate plateau exhibits a stronger dependence on $N$, thus raising doubts on the possibility of obtaining the right coexistence via integration of the equation of state across the binodal line.

In order to elucidate this point and to quantify the error, we carried out simulations of the LJ models A and B, with the method described in the previous Section. For model A, we analyzed two isothermal paths at $T=0.75$ and $T=1$. As for model B, runs were performed sequentially, moving with small steps along the integration path for $T=0.90$, with samples of $N=500,1372$, and 4000 particles in periodic cubic boxes. Even though more time-consuming, executing the runs one after the other allowed us to keep at a minimum the time needed by the structure to relax to equilibrium. For the same temperature $T=0.90$, we also simulated $1500$ model-B particles enclosed in a periodic cuboidal box with edges in the ratio of $1:1:3$.

Starting with model B in a cubic box, we first computed the chemical potential for $\rho=0.02$ and $T=1.40$ (a dilute-gas state denoted $G$) by Widom's method (at this state point the production run was 5-million cycles long). We then considered two different paths from $G$ to a liquid state $L$ ($\rho=0.75$ and $T=0.90$): one path consists of the $T=1.40$ isotherm up to $\rho=0.75$ plus a portion of the $\rho=0.75$ isochor down to $T=0.90$ (path 1); the other path (2) descends isochorically down to $T=0.90$ and then continues isothermally up to $\rho=0.75$. While path 1 circumvents the binodal line, the isothermal portion of path 2 crosses the liquid-vapor region from one side to the other. The energy and pressure equations of state for $N=1372$ along the $T=0.90$ and $T=1.40$ isotherms are shown in Fig.\,1. While $P$ is a smooth function of $\rho$ along path 1, it shows a rich structure along the $T=0.90$ isotherm, due to the occurrence of the geometric transitions mentioned above. In particular, by looking at a few system snapshots (see, {\it e.g.}, Fig.\,6 below) we were able to confirm that the liquid-drop shape changes from spherical to cylindrical to slab-like as $\rho$ increases along path 2 (the further shape transitions where the roles of vapor and liquid are inverted~\cite{MacDowell1} are not present here, likely because they would require much larger $N$ values to be resolved). The chemical potential for $T=0.90$, as either a function of pressure or density in the two-phase region, is plotted in Fig.\,2. We see that $\mu\ne\mu_{\rm coex}$ in the central part of the region, contrary to what expected for a slab configuration~\cite{MacDowell2}. We attribute this feature to a finite-size effect typical of the canonical ensemble (see Fig.\,5 below). Indeed, a similar effect is seen in the pressure, which slightly deviates from $P_{\rm coex}$ in the same density range.

The chemical potential at $L$ as computed through path 1 was found to be $-2.2942$ (with $M=5\times 10^5$ cycles in each production run). Choosing instead path 2 (now with $M=10^7$, so as to reduce the statistical error on $P$), the value of $\mu$ at $L$ turned out to be practically the same ($-2.2945$ for $N=500$ and $-2.2923$ for $N=1372$), with a residual discrepancy which we essentially ascribe to the finite step of the integration grid, causing imperfections in the spline interpolation of the pressure data especially for $N=1372$. From the crossing of liquid and vapor chemical potentials we derived the coexistence densities for $N=1372$: we found $\rho_{\rm v}=0.0451$ and $\rho_{\rm l}=0.6649$, corresponding to a coexistence pressure of $P_{\rm coex}=0.03146$, which well compare with preexisting data (see, {\it e.g.}, Table II of Ref.\,\cite{Trokhymchuk}). We have attempted to quantify the error on the chemical potential from the statistical error attached to the raw pressure data. Using standard error-propagation formulas, we estimated a precision on $\mu$ in the liquid of about one unity on the third decimal figure. This result is gratifying, since it means that pressure integration across a two-phase region is a valid tool to compute $\mu$. Obviously, choosing a path which circumvents the critical point remains the favored option to determine the chemical potential, since the pressure is then a smoother function of the density and relaxation to equilibrium is much faster.

One might think that increasing the system size, say, from $1372$ to $4000$, would entail a better estimate of the liquid $\mu$, {\it i.e.}, a better compliance with the chemical potential computed along path 1, for the obvious reason that the bulk limit is closer. In fact, this proves to be {\em false}. Using $N=4000$ with $M=4\times 10^6$ we obtained $-2.2832$, appreciably far from the value computed along path 1. This error is precisely due to the use of {\em too big} a sample. Indeed, for $N=4000$ we observed a hysteretic behavior near geometric-transition points (see Fig.\,3): the pressure values found by moving backwards in density are rather different from those registered along the forward path. Different is the case for $N=1372$, where no hysteresis was found. The occurrence of metastability for geometric transitions is clearly responsible for the failure of TI. The conclusion is that TI can be safely applied across a two-phase region, provided only that the size of the system is sufficiently {\em small} (obviously not too small, otherwise finite-size corrections will dominate).

We have also investigated the role played by the shape of the simulation box. We carried out a MC simulation of 1500 model-B particles in a cuboidal box for $T=0.90$ (with $M=4\times 10^6$ cycles in each production run). In Fig.\,4 we make a comparison of the equations of state for this system with those for $1372$ particles in a cubic box. In the two-phase region and in spite of the similar sizes, we see that the pressure of the two systems are largely different, due arguably to an explicit dependence of the geometric-transition thresholds on the box aspect ratio. Note, in particular, that in the elongated-box case the spherical ``phase'' is washed out (as was checked by inspection of many system configurations) whereas the cylindrical ``phase'' is greatly reduced in extent. Notwithstanding the difference in pressure between the two systems, their coexistence parameters are very close: for $N=1500$ we found $P_{\rm coex}=0.03149,\mu_{\rm coex}=-3.0523,\rho_{\rm v}=0.0452$, and $\rho_{\rm l}=0.6649$; finally, the chemical potential at point $L$ was found to be $-2.2912$, again close to that computed along path 1.

The top panel of Fig.\,5 shows the pressure equation of state near the center of the liquid-vapor region for model-B systems of various sizes. We see that, for $\rho\simeq(\rho_{\rm v}+\rho_{\rm l})/2$, the difference $P_{\rm coex}-P(\rho)$ becomes smaller as the size of the system grows. The system in the elongated box is an exception, in that the discrepancy is smaller than for a sample of comparable size ($N=1372$) but enclosed in a cubic box. The same effect is seen in the chemical potential (bottom panel): the value of $\mu$ at the center of the two-phase region is closer to $\mu_{\rm coex}$ for $N=1500$ (elongated box) than for $N=1372$ (cubic box).

Finally we looked at model A for two different subcritical temperatures, $T=0.75$ and $T=1$. We made no attempt to perform sequential simulations for this system but rather ran the MD code for each state point independently, using an initial fcc configuration for every density. A rapid glance at Fig.\,6 shows that a further pressure ``plateau'' now shows off in a $\rho$ range between those relative to cylindrical- and slab-shaped liquid drops. By looking at the typical system configuration in this density range it appears that the liquid gives rise to an unusual, brand-new arrangement: a slab with a circular hole inside (see an example in Fig.\,7). The same evidence was found at the higher temperature $T=1$, though the extra plateau is now narrower. However, the extra plateau for $T=1$ soon disappeared when we doubled the length of the MD trajectory while nothing similar happened for $T=0.75$. We argue that the hollow-slab structure of the liquid drop is a manifestation of a stable (or nearly stable) heterogeneous ``phase'' of the system, at least for sufficiently low temperatures. This structure did not emerge in our model-B simulations likely because $T=0.90$ is not too low a temperature, or for the simple reason that sequentially-generated configurations unavoidably bear some memory of the structure of the system in the previous run performed with a slightly smaller density, a bias not present when runs are performed in parallel.

\subsection{The solid-liquid transition of the LJ system}

We have seen that a straightforward $NVT$ simulation is able to reproduce the subtle structure of the LJ fluid inside the liquid-vapor region. In addition, TI works correctly across the two-phase region, at least provided the system is not too big. This happens thanks to the fast relaxation of the system to equilibrium: in spite of the non-zero cost of interface formation, the density of the heterogeneous fluid is not as large as to prevent particles from finding their place in the overall system architecture in a time affordable by an MD or MC simulation. Different is the case of a dense fluid approaching solidification at constant temperature. In this case relaxation times are so long that a simulation with local moves simply fails to notice the existence of a more stable crystalline phase and the fluid then goes metastable. A noteworthy exception is Ref.\,\cite{Bernard} where, thanks to the reduced system dimensionality and, especially, to a smart simulation method, one observes a few geometric transitions also in the solid-liquid region. In Ref.\,\cite{Statt}, the existence of shape transitions was instead implicit in the kind of heterogeneity emerging out of a very long run carried out at selected densities within the solid-liquid region on a system initially prepared in a liquid configuration with a solid droplet inside. Hence, the right question one should ask is whether in an {\em ordinary} simulation things can be so arranged that the error made in estimating coexistence parameters by TI can be kept at a minimum.

We simulated model A for various temperatures ($T=0.75,1.15,1.35,2.00,2.74$) by preparing 1372 particles in a perfect fcc configuration and then letting them evolve for a fixed density (as is well known, for sufficiently high temperatures the stable LJ crystal is fcc rather than hexagonal close packed, see, {\it e.g.}, Ref.\,\cite{Travesset}). We wanted to check down to what density the crystal withstands the thermal motion of the particles for long without melting. For instance, for $T=1.15$ (1.35) we know from Ref.\,\cite{Hansen} that the coexistence densities are $\rho_{\rm l}=0.936$ (0.964) and $\rho_{\rm s}=1.024$ (1.053). We report the final system pressure as a function of density in Fig.\,8. We see that the undercompression of the crystal is indeed small, but the values found for the pressure in the coexistence region are admittedly wrong. Upon interpolating these data with spline functions, we found $\rho_{\rm l}=0.907$ (0.934) and $\rho_{\rm s}=0.992$ (1.000), about 8\% (4\%) away from the known thresholds.

Moving to model B, Pedersen has recently computed the transition boundaries for $T=0.90$ (1.40)\,\cite{Pedersen}, finding $\rho_{\rm l}=0.903$ (0.988) and $\rho_{\rm s}=0.989$ (1.061), for a coexistence pressure of $P_{\rm coex}=3.514$ (11.181). We first checked our MC code and method against these benchmarks by computing, for $N=1372$ particles in a cubic box with PBC, the ``exact'' Helmholtz free energy of the fcc crystal for $\rho=1.20$ and $T=0.50$ by the Frenkel-Ladd method. Using this state for reference, TI allows one to obtain the chemical potential of the fcc crystal in any other state where this phase is stable or metastable. Upon comparing the crystal $\mu$ with the fluid chemical potential, we located the phase transition at $P_{\rm coex}=3.459$ (11.066) for $T=0.90$ (1.40), whence obtaining $\rho_{\rm l}=0.901$ (0.987) and $\rho_{\rm s}=0.988$ (1.060), which are extremely close to Pedersen's values (for $T=0.90$, the chemical potentials of liquid and solid are plotted throughout the relevant pressure range in Fig.\,9).

We reported our energy and pressure data for the isothermally compressed fluid in Fig.\,10, where we see that the fluid goes deeply metastable at both temperatures, until it abruptly transforms at a certain density into a (defective) crystal (both the energy and pressure of the solid resulting from the decay of the metastable fluid are larger than those of the fcc crystal). Also the fcc crystal became metastable but the undercompression of the crystal is a moderate effect compared to the overcompression of the fluid. In Fig.\,11 we plotted the elastic constants (see for example Ref.\,\cite{Prestipino2}) and Steinhardt order parameters (OPs)\,\cite{Steinhardt} of the system along the fcc branch. We see that both the elastic moduli and orientational OPs vanish at the ultimate metastability threshold of the crystal rather than at the melting density. Hence, there is no hope of getting $\rho_{\rm s}$ (and $P_{\rm coex}$ as well) from the vanishing of the crystal OPs.

If we were to estimate the coexistence parameters from the integration of $P(\rho)$ along the fluid branch we would obtain a pressure which, for example, at $T=1.40$ is wrong by as much as 80\%. This wrong estimate of $P$ even slightly worsens should we go up in size from 1372 to 4000 (for a $M$ value of equal magnitude). That is, notwithstanding there are more opportunities in a larger system for the solid to nucleate spontaneously, the more pronounced inertia of the bigger liquid to convert into a solid eventually prevails. This evidence suggests that a better strategy to reduce the error of $P_{\rm coex}$ is to {\em decrease} (rather than increase) the system size. As shown in Fig.\,12, this expedient actually works: the decay of the metastable fluid into a solid occurs for a density which is systematically lower the smaller the system. The best result ever (that is, $P_{\rm coex}$ overestimated by ``only'' 16\% for $T=1.40$) was obtained for $N=108$, a system so small that for high densities we had to include energy and pressure tail corrections in terms of a radial distribution function computed beyond half box edge, up to $2.5\sigma$ (see Ref.\,\cite{Theodorou} for details). Looking retrospectively, we conclude that PBC stabilize the crystalline phase in a small system more effectively than a larger availability of nuclei centers would do in a large system.

A still better estimate of coexistence thresholds is obtained if, as soon as the metastable fluid has decayed into solid, we start following the defective-solid branch backwards in density until the system re-melts (to accomplish this purpose, it is not necessary to know the crystalline structure in advance). We expect a small-size crystal with imperfections to melt upon isothermal expansion very near the true melting point; this is why the minimum pressure of this solid is a rather good estimate of $P_{\rm coex}$: for model B ($N=256$) we predicted $P_{\rm coex}=3.099$ (10.532) for $T=0.90$ (1.40), both estimates being within 10\% percent of the exact values.

\section{Conclusions}

A cheap way to work out the entire phase diagram of a model fluid is by following the isothermal evolution of the pressure $P$ as a function of density $\rho$, any loop in $P(\rho)$ being the hallmark of the crossing of a two-phase region (see for instance Ref.\,\cite{Abramo,Hus,Dudalov}). However, traditional simulation methods usually fail to attain thermal equilibrium near transition points and the question is whether it is possible to obtain accurate transition thresholds from an equation of state which is only approximate in the coexistence regions.

To this aim we have studied the Lennard-Jones model as its behavior is paradigmatic for most fluids. We first verified that the liquid chemical potential is indeed exactly recovered through pressure integration along a path crossing the liquid-vapor coexistence region, at least unless the sample is too big (for a too large system the chemical-potential estimate is far from perfect because of pressure hysteresis close to morphological transitions of the liquid drop). Conversely, the same confidence cannot be placed in the chemical potential computed along a fluid branch running across the solid-liquid region. It is a fact that neither molecular dynamics nor Metropolis Monte Carlo simulation can achieve equilibration in the solid-liquid region in a sufficiently short time. If we anyway decide to avoid using smarter simulation methods, the most we can do is to employ systems of moderate size since too large systems incur a more serious form of metastability. We found that a decent estimate of the transition pressure (accurate to within 5-10\%) can be obtained by isothermally expanding until melting the defective solid eventually emerged from the freezing of the metastable liquid. Clearly, the error remains and it is at the least problematic to infer reliable melting and freezing densities from a non-monotonic pressure by plain thermodynamic integration, especially when the phase diagram is suspected to host many crystalline phases of nearly equal stability. A more systematic analysis of the merits and pitfalls of pressure integration, in comparison with other heuristic methods (along the lines of what done in, {\it e.g.}, Refs.\,\cite{Zhang,Prestipino3,Saija}), is deferred to a future study.

\acknowledgements
MCA, CC, DC, and G. Muna\`o acknowledge support from PRIN-MIUR 2010-2011 Project.

\newpage
\begin{center}
\noindent FIGURE CAPTIONS
\end{center}

Fig.\,1:
LJ model B, energy and pressure equations of state for $1372$ particles in a cubic box with PBC (left: $T=0.90$; right: $T=1.40$). Here, only the range from low to moderate densities is examined. Statistical averages were computed over $M=10^7$ MC cycles. The three pressure plateaus observed in the left-bottom panel correspond, in order of increasing density, to a spherical, a cylindrical, and a slab-like droplet of liquid immersed in a vapor (see next Fig.\,6). The vertical lines mark the coexistence densities of vapor and liquid, $\rho_{\rm v}$ and $\rho_{\rm l}$; the full horizontal line is located at the coexistence pressure, $P_{\rm coex}$.

Fig.\,2:
LJ model B, chemical potential of 1372 particles in a cubic box with PBC, plotted as a function of either pressure or density for $T=0.90$ across the liquid-vapor region. While in the top panel we observe the cusp of $\mu(P)$ at the transition, the three plateaus corresponding to spherical, cylindrical, and slab-like liquid droplets are seen in the bottom panel. The vertical lines stay at the coexistence densities of vapor and liquid, whereas the horizontal line marks the coexistence chemical potential $\mu_{\rm coex}=-3.0530$.

Fig.\,3:
LJ model B, energy and pressure equations of state at low density for $T=0.90$: $N=1372$ (black open dots) and $N=4000$ (red open squares). For every state point, averages were computed over $M=4\times 10^6$ MC cycles. We show the effect of retracing the $e$ and $P$ curves backwards in density (full dots and squares): while nothing particular happens for $N=1372$ ({\it i.e.}, the system travels the thermodynamic path reversibly), hysteresis is found close to geometric-transition thresholds for $N=4000$ (this effect is less visible for the energy but nonetheless present). The vertical line marks the position of $\rho_{\rm v}$ whereas the full horizontal line stays at the level of $P_{\rm coex}$.

Fig.\,4:
LJ model B, energy and pressure equations of state for $T=0.90$: $N=1372$ particles enclosed in a cubic box with PBC (black dots) and $N=1500$ particles enclosed in a cuboidal box with PBC (blue squares). The statistical errors affecting the data are smaller than the size of the symbols. The vertical lines mark the positions of $\rho_{\rm v}$ and $\rho_{\rm l}$. A full horizontal line has been drawn at $P_{\rm coex}$.

Fig.\,5:
LJ model B, pressure (top) and chemical-potential equation of state (bottom) near the center of the liquid-vapor region for systems of various sizes: $N=500$ (green diamonds), $N=1372$ (black dots), $N=4000$ (red triangles), and $N=1500$ (blue squares, elongated box). The arrow in the top panel marks the center of the two-phase region. The horizontal lines are drawn at the coexistence pressure (top) or chemical potential (bottom) for $N=1372$. The vertical lines mark the vapor coexistence density $\rho_{\rm v}$.

Fig.\,6:
LJ model A, pressure equation of state for $T=0.75$ (only the range from low to moderate densities is shown). In addition to the three expected ``plateaus'' (see Fig.\,1) we see another one, marked in the figure, where the liquid drop has the shape of a punched slab (see an example in the next Fig.\,7). The miniatures show typical system configurations in the regions indicated by the arrows. Particles (which here were given a diameter of $\sigma$) have been colored differently, according to the number $n_{\rm NN}$ of nearest neighbors (NN) of each (two particles are said to be NN of each other if they stay within a distance $r_{\rm min}$, which is where the radial distribution function of the liquid at coexistence attains its first non-zero minimum). The color code is as follows: $n_{\rm NN}=4$ or 5, yellow; $n_{\rm NN}=6$ or 7, magenta; $n_{\rm NN}=8$ or 9, green; $n_{\rm NN}=10$ or 11, cyan; $n_{\rm NN}\ge 12$, blue (particles with $n_{\rm NN}\le 3$ were not plotted).

Fig.\,7:
LJ model A, a snapshot of the system configuration for $\rho=0.262$ and $T=0.75$. The liquid drop has the shape of a slab with a circular hole. The color of each particle has been decided on the basis of the number of NN, as explained in Fig.\,6 caption. White particles are those with $n_{\rm NN}\le 3$.

Fig.\,8:
LJ model A, pressure equation of state for $T=1.15$ (blue dots) and $T=1.35$ (red squares) in the high-density region, obtained by performing a long MD simulation at each density from an initial perfectly-ordered fcc configuration ($N=1372$). For, {\it e.g.}, $T=1.15$, at the end of the simulation the crystal still keeps its structure for $\rho\gtrsim 0.95$ while it has eventually melted for smaller $\rho$ values.

Fig.\,9:
LJ model B, chemical potential vs. pressure for $N=1372$ and $T=0.90$. The phase transitions are evidenced in the two cusps of $\mu(P)$, at $(0.03146,-3.0530)$ (liquid-vapor transition) and at $(3.4589,1.1527)$ (solid-liquid transition). A magnification of the transition regions is shown in the insets (right bottom inset, liquid-vapor transition; left top inset, solid-liquid transition). The thin horizontal and vertical lines in the insets mark the location of the transition points.

Fig.\,10:
LJ model B, energy and pressure equations of state at high density for $T=0.90$ (left) and $T=1.40$ (right) along the fluid (black open dots) and the fcc-crystal branch (blue full dots), for $N=1372$. By exact free-energy calculations the transition for $T=0.90$ (1.40) is located at $P_{\rm coex}=3.461$ (11.066), marked by a horizontal line, in perfect agreement with the thresholds reported in Table I of Ref.\,\cite{Pedersen}. The vertical lines represent the exact melting and freezing densities. We have also reported (as red open dots) the energy and pressure values obtained by moving backwards in density from the last point reached along the fluid branch.

Fig.\,11:
LJ model B, cubic elastic constants and Steinhardt parameters along the fcc-crystal branch for $T=0.90$ (left) and $T=1.40$ (right). The three elastic constants were computed by the formulas derived in Ref.\,\cite{Farago} ($c_{11}$, red crosses; $c_{12}$, blue squares; $c_{44}$, black dots). The orientational OPs, $q_4$ (squares) and $q_6$ (dots), were obtained by the method described in Ref.\,\cite{tenWolde}. The vertical lines represent the exact melting and freezing densities. Both $c_{44}$ and the orientational OPs are found to vanish exactly where the metastable crystal melts (see Fig.\,10).

Fig.\,12:
LJ model B, energy and pressure equations of state at high density (fluid branch, open symbols) for systems of various sizes: $N=108$ (blue triangles), 256 (inverted triangles), 500 (diamonds), and 4000 (red squares). The data points for $N=108$ and $N=4000$ were joined by straight-line segments to guide the eye. The energy and pressure for a 1500-particle system in a cuboidal box are also shown (crosses and dotted lines). Left: $T=0.90$; right: $T=1.40$. Full symbols refer to data points for $N=256$ obtained by following the solidified system backwards in density.

\newpage
%
%
\begin{figure}
\centering
\includegraphics[width=13cm]{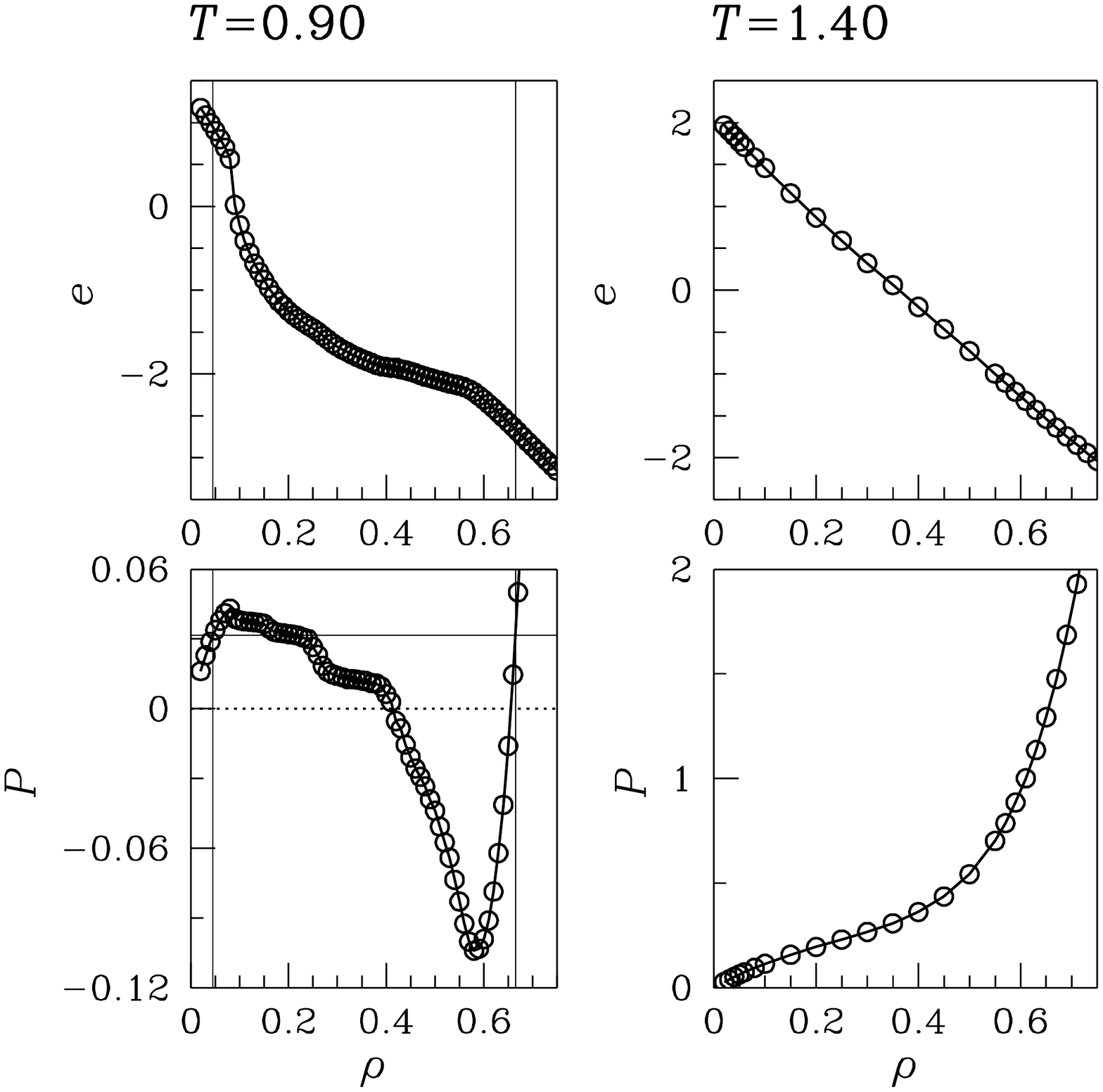}
\caption{
}
\label{fig1}
\end{figure}

%
%
\begin{figure}
\centering
\includegraphics[width=13cm]{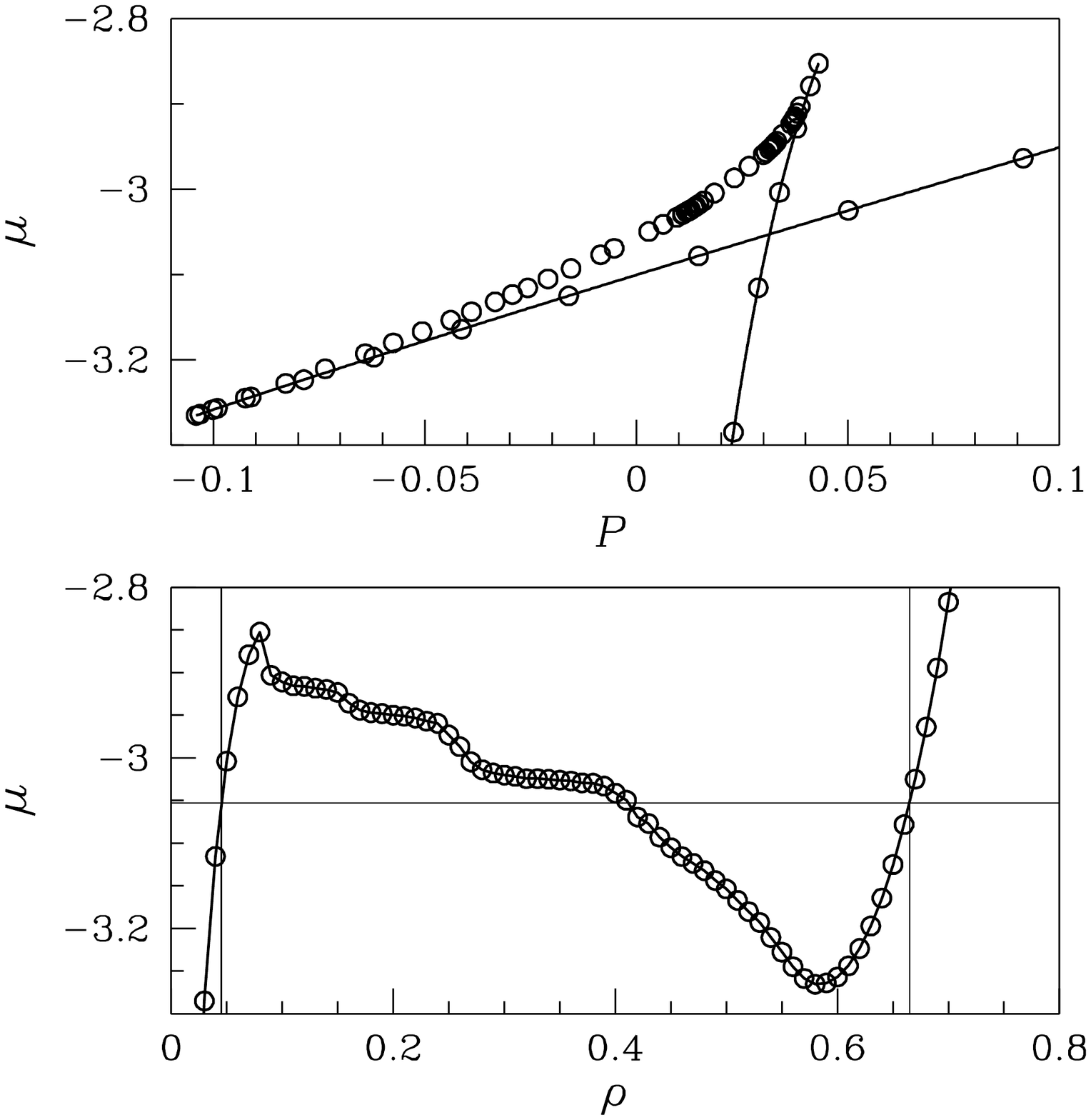}
\caption{
}
\label{fig2}
\end{figure}

%
%
\begin{figure}
\centering
\includegraphics[width=13cm]{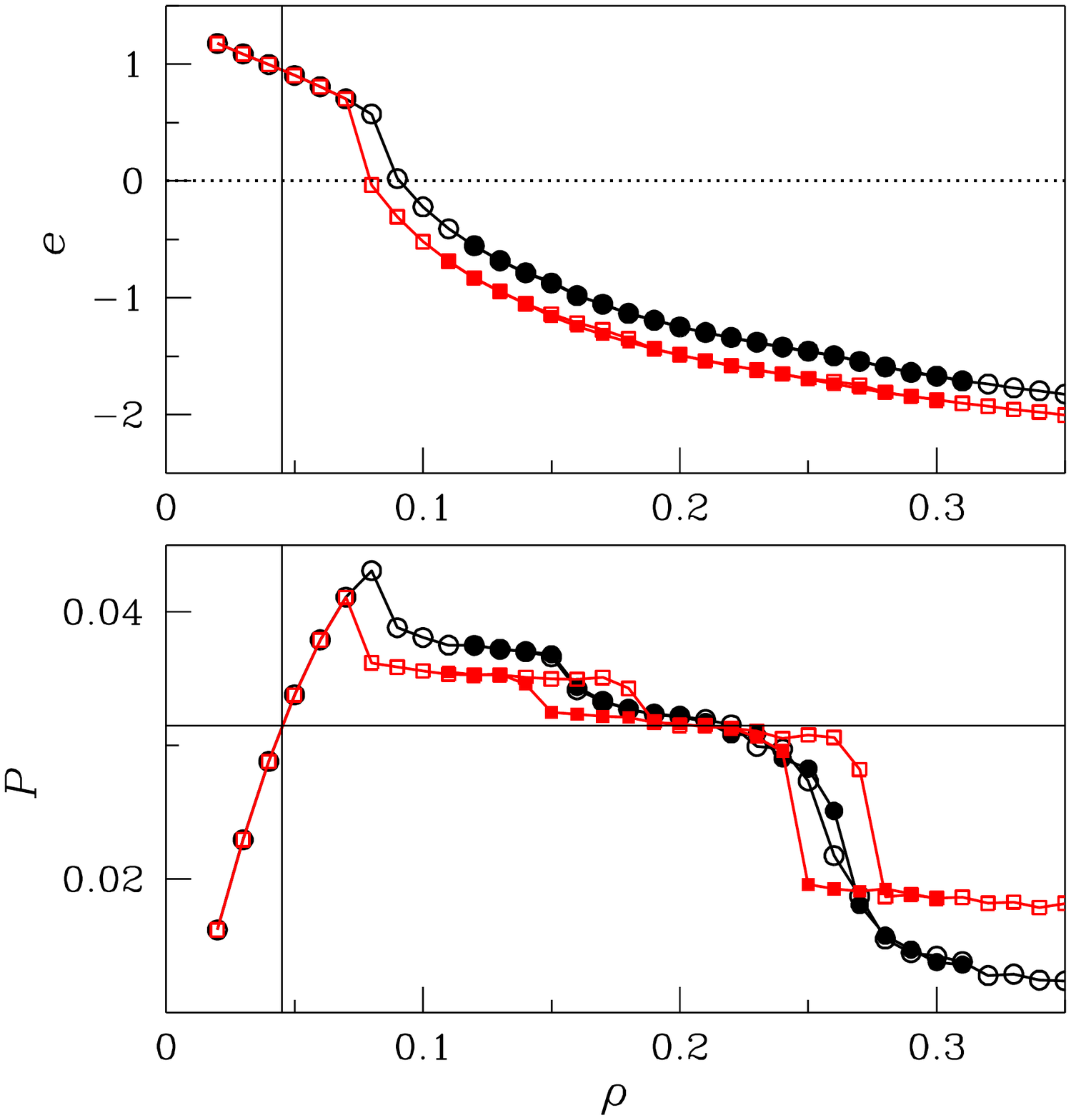}
\caption{
}
\label{fig3}
\end{figure}

%
%
\begin{figure}
\centering
\includegraphics[width=13cm]{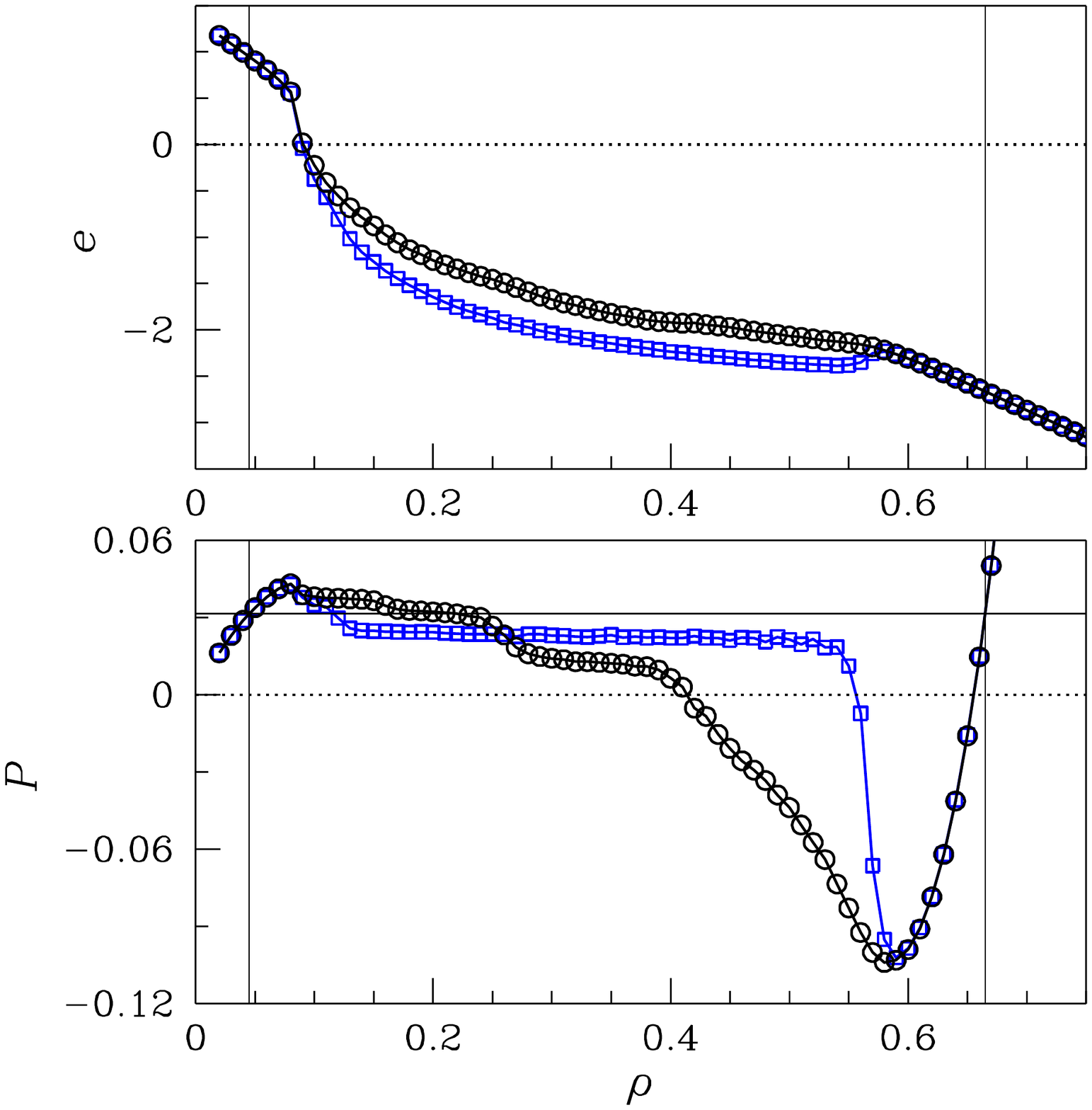}
\caption{
}
\label{fig4}
\end{figure}

%
%
\begin{figure}
\centering
\includegraphics[width=13cm]{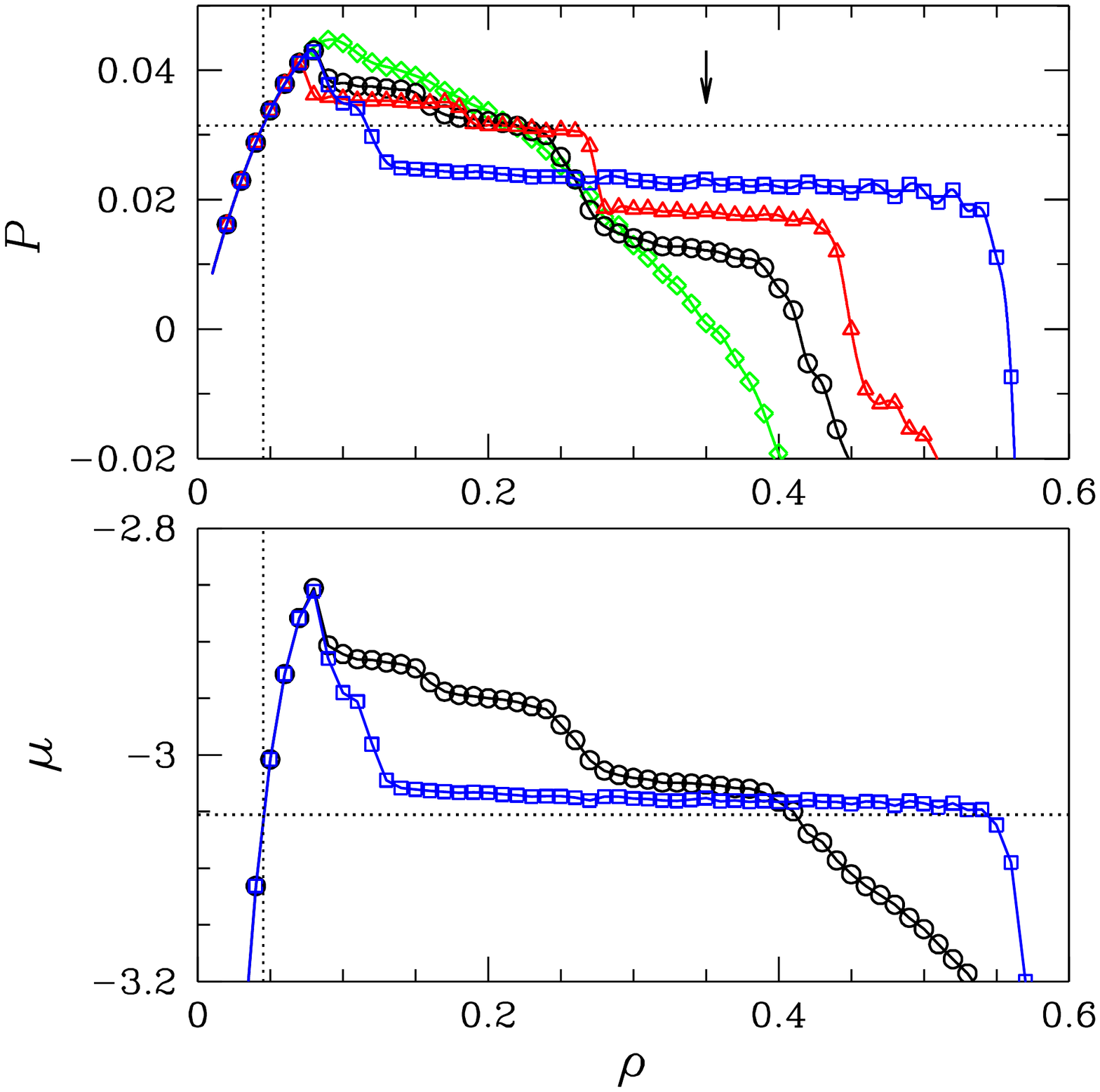}
\caption{
}
\label{fig5}
\end{figure}

%
%
\begin{figure}
\centering
\includegraphics[width=13cm]{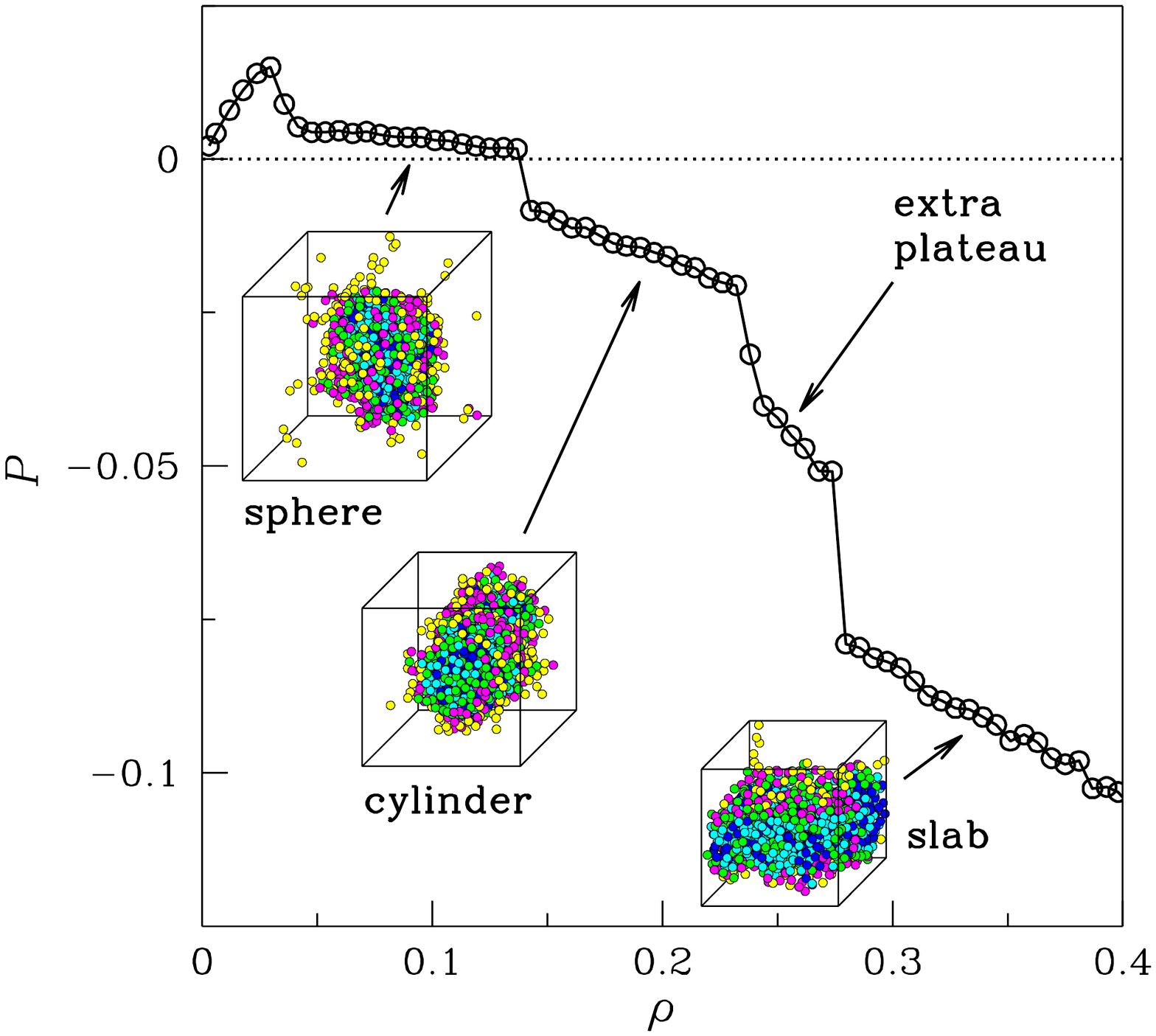}
\caption{
}
\label{fig6}
\end{figure}

%
%
\begin{figure}
\centering
\includegraphics[width=13cm]{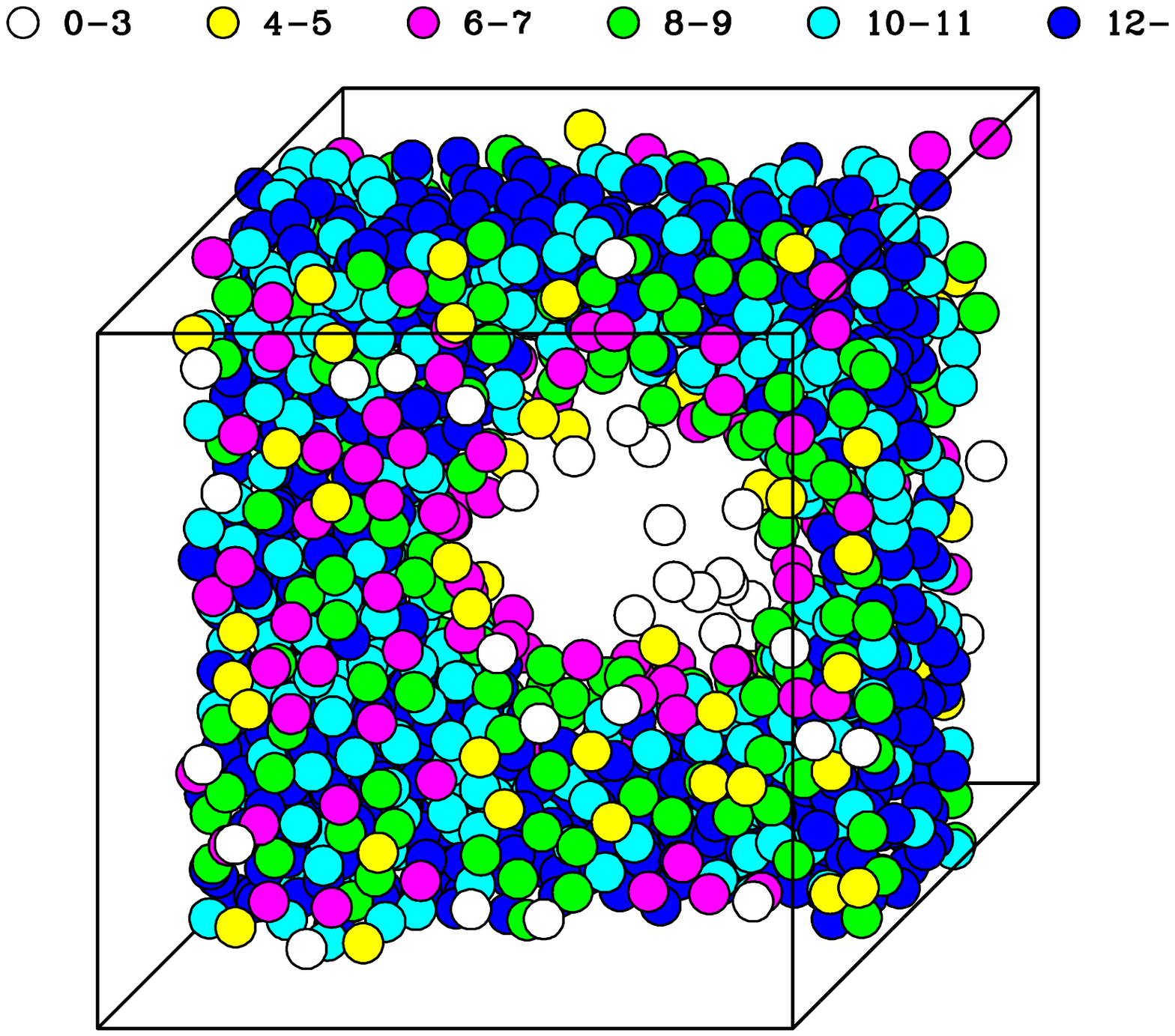}
\caption{
}
\label{fig7}
\end{figure}

%
%
\begin{figure}
\centering
\includegraphics[width=13cm]{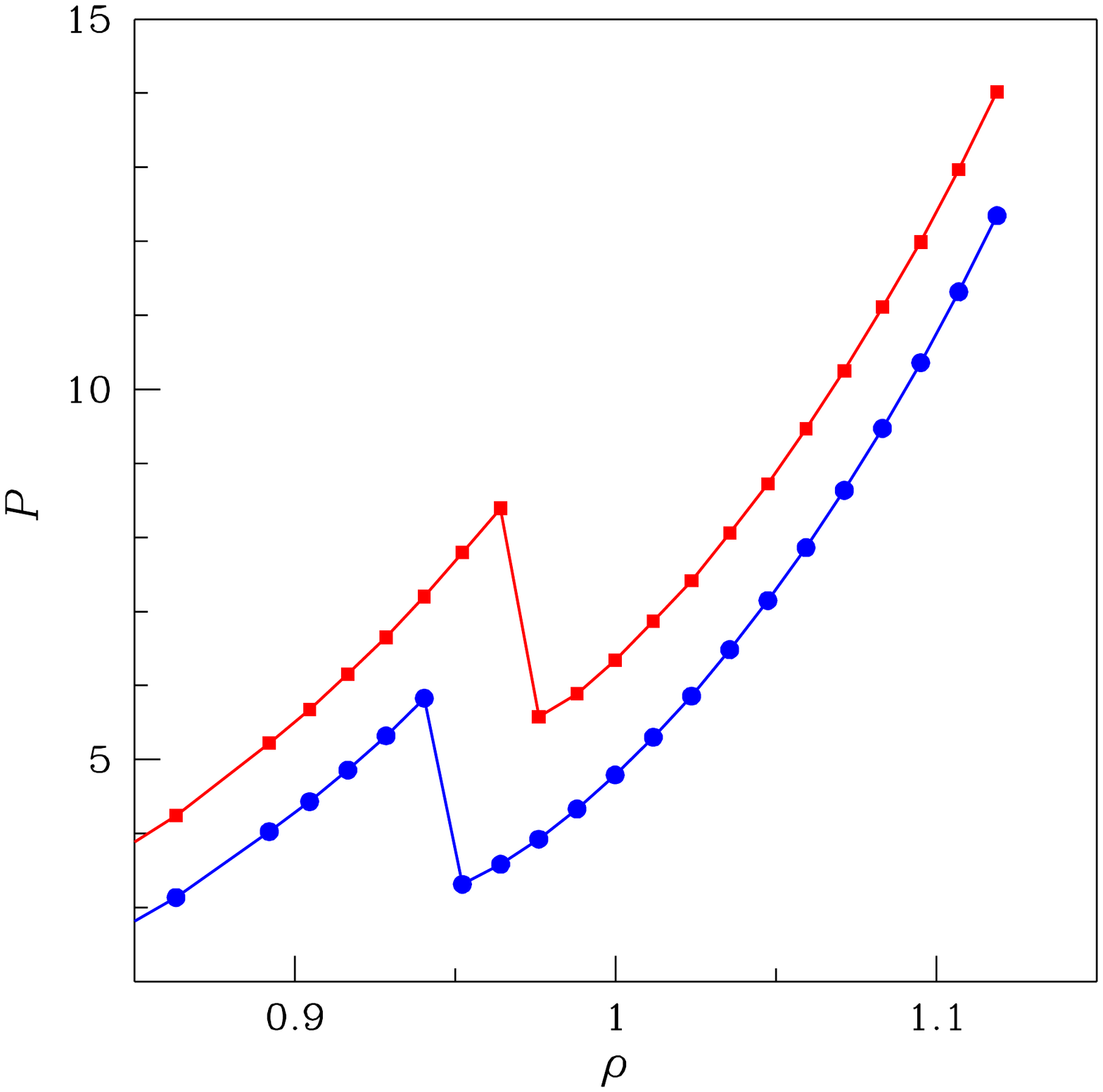}
\caption{
}
\label{fig8}
\end{figure}

%
%
\begin{figure}
\centering
\includegraphics[width=13cm]{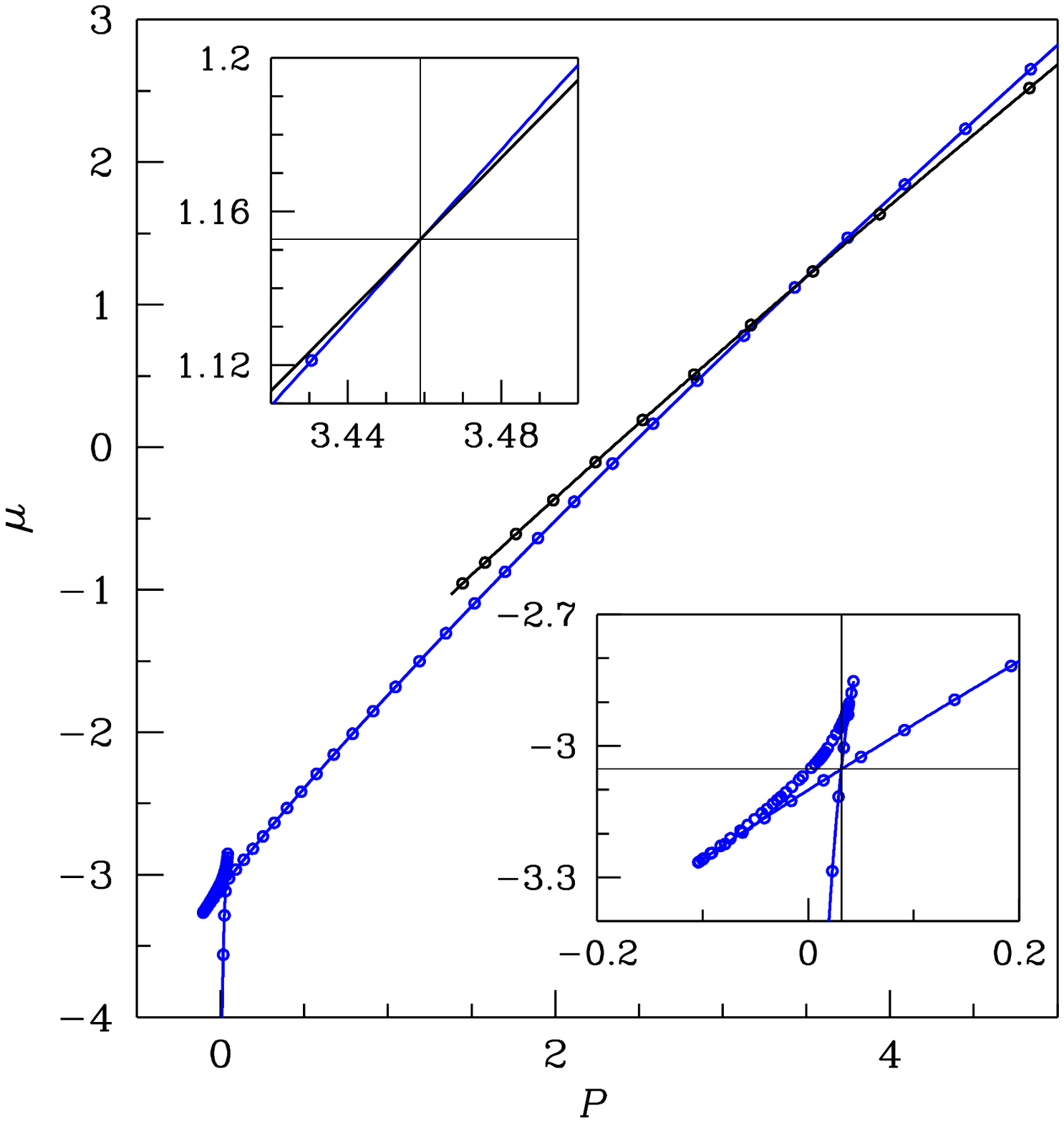}
\caption{
}
\label{fig9}
\end{figure}

%
%
\begin{figure}
\centering
\includegraphics[width=13cm]{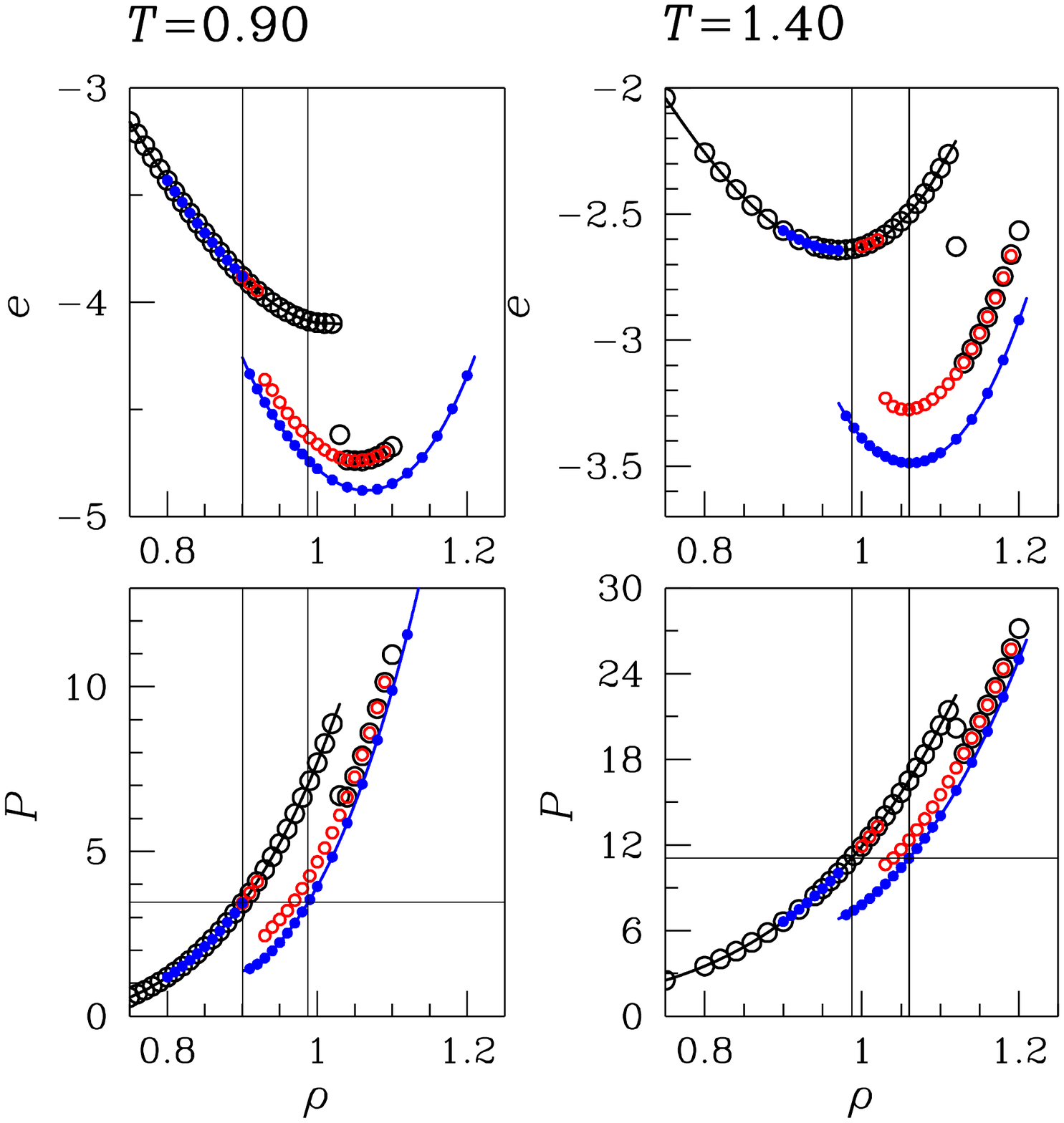}
\caption{
}
\label{fig10}
\end{figure}

%
%
\begin{figure}
\centering
\includegraphics[width=13cm]{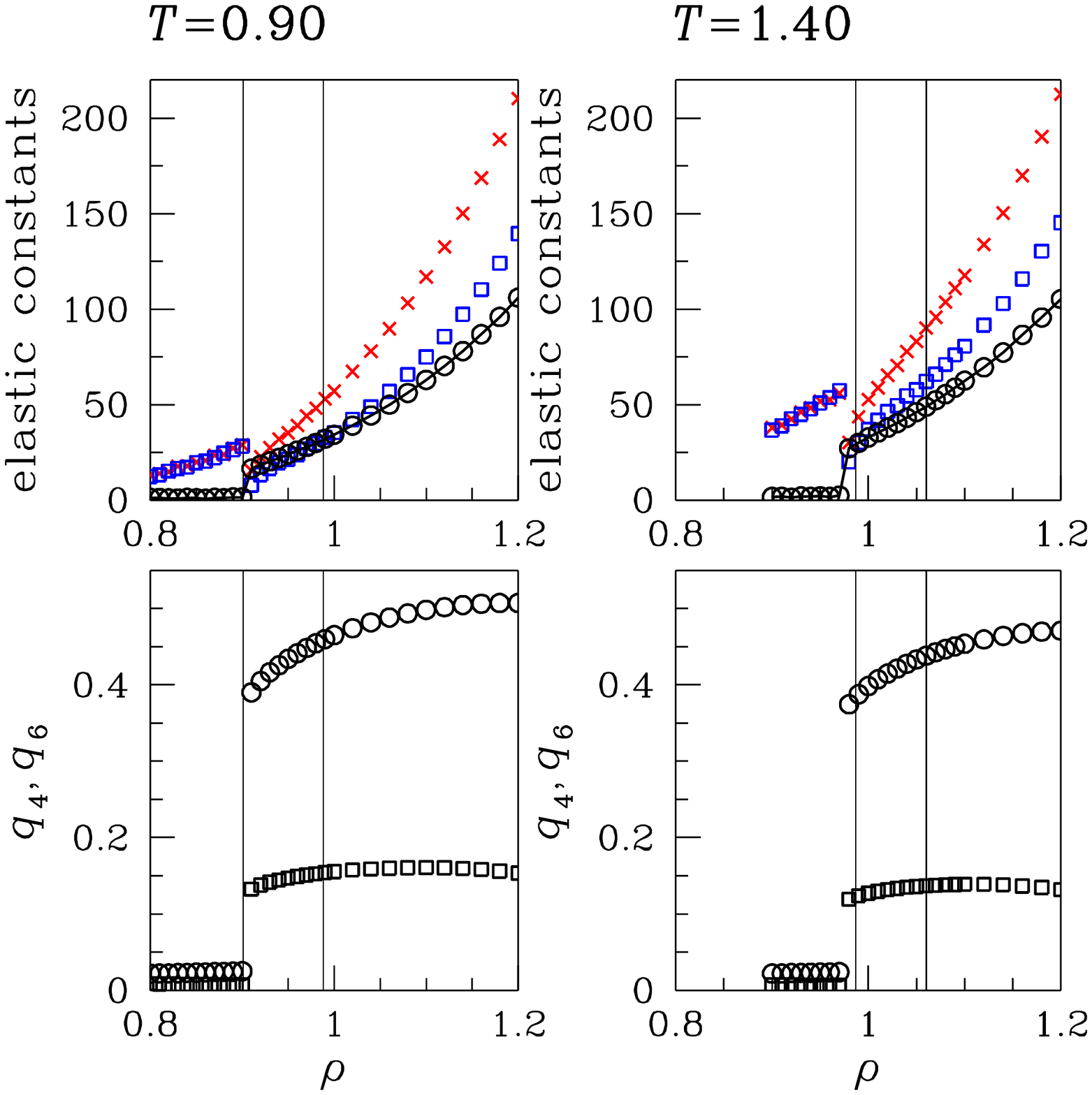}
\caption{
}
\label{fig11}
\end{figure}

%
%
\begin{figure}
\centering
\includegraphics[width=13cm]{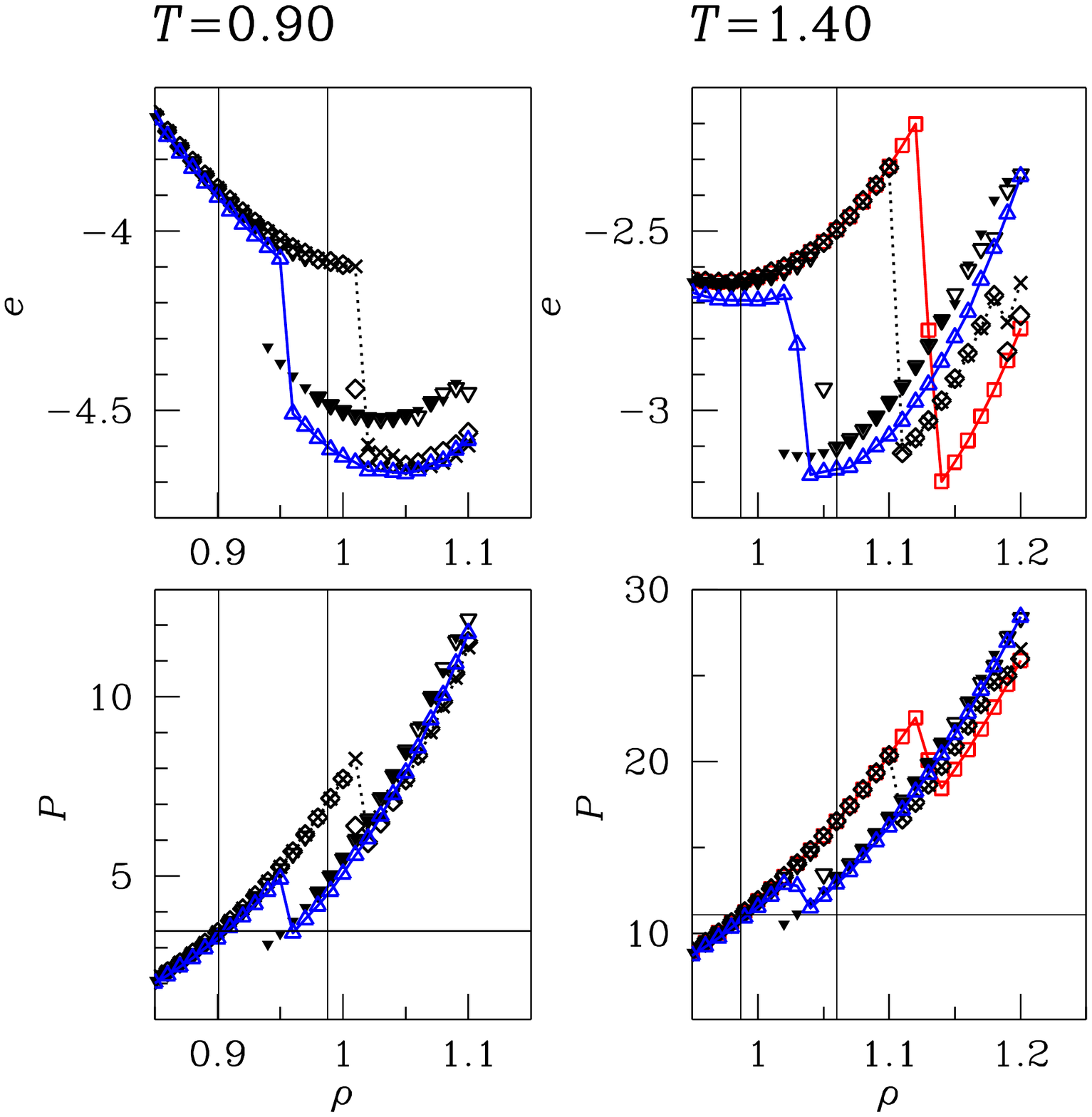}
\caption{
}
\label{fig12}
\end{figure}
\end{document}